\documentclass[10pt, a4paper, twocolumn]{article}
\usepackage{xurl}
\usepackage[super,comma,sort&compress]{natbib}
\usepackage{graphicx}
\usepackage{amsmath,amssymb,amsfonts,bbm}
\usepackage{soul,color}
\usepackage{authblk}
\usepackage[margin={1.5cm,2.5cm}]{geometry}
\usepackage{abstract}

\date{}

\title{\vspace{-30pt}Enriched Music Representations with Multiple Cross-modal Contrastive Learning}

\author{Andres Ferraro$^{*}$, Xavier Favory\thanks{A. Ferraro, X. Favory and D. Bogdanov are with Music Technology Group, Universitat Pompeu Fabra, Spain, email \{firstname.lastname\}@upf.edu}, Konstantinos Drossos\thanks{K. Drossos is with Audio Research Group, Tampere University, Finland, email konstantinos.drossos@tuni.fi}, Yuntae Kim\thanks{Y. Kim is with Kakao Corp, Korea, email daniel.py@kakaocorp.com} and Dmitry Bogdanov$^{*}$
\thanks{Thanks to Tuomas Virtanen, Soohyeon Lee and Biho Kim for their valuable feedback. This work was partially supported by Kakao Corp. K. Drossos was partially supported by the European Union’s Horizon 2020 research and innovation programme under grant agreement No 957337, project MARVEL}
}

\begin{document}
\maketitle
\begin{abstract}
Modeling various aspects that make a music piece unique is a challenging task, requiring the combination of multiple sources of information. 
Deep learning is commonly used to obtain representations using various sources of information, such as the audio, interactions between users and songs, or associated genre metadata. 
Recently, contrastive learning has led to representations that generalize better compared to traditional supervised methods.
In this paper, we present a novel approach that combines multiple types of information related to music using cross-modal contrastive learning, allowing us to learn an audio feature from heterogeneous data simultaneously.
We align the latent representations obtained from playlists-track interactions, genre metadata, and the tracks' 
audio,
by maximizing the agreement between these modality representations using a contrastive loss. 
We evaluate our approach in three tasks, namely, genre classification, playlist continuation and automatic tagging.
We compare the performances with a baseline audio-based CNN trained to predict these modalities.
We also
study the importance of including multiple sources of information when training our embedding model. The results suggest that the 
proposed method outperforms the baseline in all the three downstream tasks and achieves comparable performance to the state-of-the-art. 
\end{abstract}
%
%
%


\section{Introduction and Related Work}

There are multiple sources and types of information related to the music that can be used for different applications. For example, using audio features showed better performance for predicting musical genres compared to using users' listening data~\cite{won2020multimodal}. On the other hand, the latter performed better
on music recommendation~\cite{celma2008new} and mood prediction~\cite{korzeniowski2020mood}. Having a numerical feature representation that combines all the relevant information of a song would allow creating better automatic tools that solve problems such as genre prediction, mood estimation and music recommendation.
%

Advances of deep learning in the past years enabled to improve the performance on multiple tasks by combining different types of data. For example, Oramas et al. \cite{oramas2018multimodal} propose a multi-modal approach combining text, audio, and images for music auto-tagging and Suris et al. \cite{suris2018cross} propose a method to combine audio-visual embeddings for cross-modal retrieval.

Deep learning allows learning representations mapping from different input data to an embedding space that can be used for multiple downstream tasks~\cite{radford2015unsupervised}. 
The most common approach for representation learning in the music domain is to train a audio-based classifier to predict some music aspects such as genre, mood, or instrument and then use the pre-trained model to extract embeddings that could be used in different tasks. Alonso et al.~\cite{alonsoembeddings} compare different pre-trained architectures for 
predicting multiple aspects of a song such as danceability, mood, gender and timbre, showing the generalization capabilities of these pre-trained models.  
Alternative methods in the field of deep metric learning 
recently shown a better performance across multiple downstream tasks compared to the approach of pre-training classification models~\cite{zhai2018classification, leemetric}, demonstrating the great potential of deep metric learning for generalizing to a larger diversity of tasks. 
%

Contrastive learning has gained popularity in the last years~\cite{le2020contrastive}. These approaches allow to learn representation by employing a metric learning objective, contrasting similar and dissimilar items. The similar examples are referred as positive examples and the dissimilar are referred as negative examples. 
Approaches based on triplet loss~\cite{weinberger2009distance} 
require to define triplets composed of an anchor, a positive and a negative example. Triplet loss was recently applied in the music domain for  retrieval~\cite{won2020multimodal} and zero-shot learning~\cite{choi2019zero}. However, the strategy for sampling the triplets is crucial to the learning process and can require significant effort. 
There are other losses that instead of defining triplets rely on the comparison of paired examples such as \emph{infoNCE} \cite{oord2018representation} and \emph{NT-Xent} \cite{chen2020simple}. 
They have the advantage of involving all the data points within a mini-batch when training without requiring to define a specific strategy for sampling the training examples. Employing these contrastive loss functions in a self-supervised way has led to powerful image~\cite{chen2020simple}, sound~\cite{fonseca2020unsupervised} and music audio~\cite{saeed2020contrastive} representations learned without the need for annotated data. 
Contrastive learning was also applied in a supervised way~\cite{favory2020coala,khosla2020supervised} with a cross-modal approach using sound (audio) information and associated text metadata in order to learn semantically enriched audio features. The learned features achieve competitive performance in urban sound event and musical instrument recognition~\cite{favory2020coala}. 

The works mentioned above suggest that methods based on contrastive learning
have the potential to exploit different types of data which is promising for improving the performance of deep audio embeddings for a large diversity of tasks.
However, to our knowledge, there is no work that focuses on leveraging multiple modalities through contrastive learning in order to learn rich musical audio features. 
This motivates us to investigate approaches that take advantage of different types of music-related information (i.e. audio, genre, and playlists) to obtain representations from the audio that can perform well in multiple downstream tasks such as music genre classification, automatic playlist continuation, and music automatic tagging. Our results show that the proposed contrastive learning approach reaches performance comparable to the state-of-art and outperforms models pre-trained for classification or regression based on the musical aspect.
%
%
%

Our contributions are as follows: 
i) We propose an updated audio encoder optimized for the music domain 
based on the approach proposed by Favory et al.~\cite{favory2020learning, favory2020coala}.
ii) We use the alignment of multi-modal data for exploiting the semantic metadata and collaborative filtering information. iii) We evaluate the obtained representations in three downstream tasks using different datasets comparing with other common approaches based on pre-training for classification or regression. iv) We also include an ablation study by comparing the performance of each source of information independently, which allows us to understand the importance of the different parts of our model.~\footnote{We provide the code to reproduce this work and the pre-trained models: \url{https://github.com/andrebola/contrastive-mir-learning}}
\section{Proposed method}\label{sec:method}
\begin{figure}[!t]
    \centering
    \includegraphics[width=.35\textwidth,trim={0.05cm 0 0.05cm 0.05cm},clip]{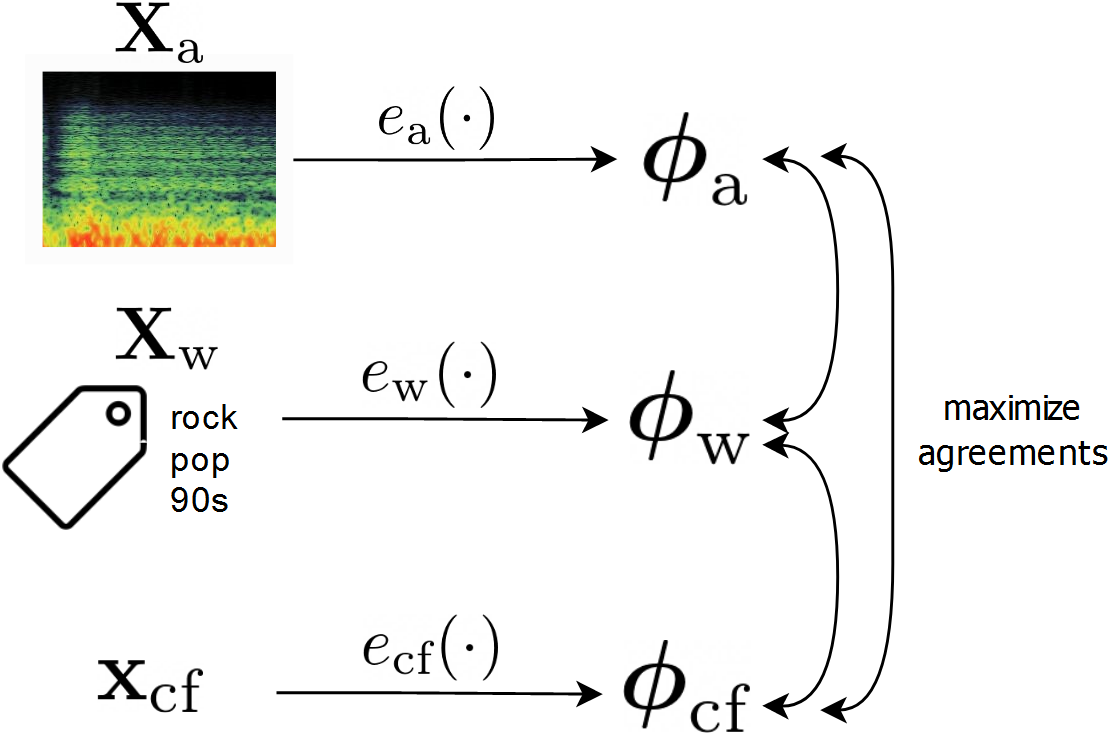}
    \caption{Diagram with architecture of the method}
    \label{fig:arch}
\end{figure}

Our method employs the encoders $e_{\text{a}}(\cdot)$, $e_{\text{w}}(\cdot)$, and $e_{\text{cf}}(\cdot)$, encoding audio and embeddings of musical genres and music playlist information, respectively, and a dataset $\mathbb{D}=\{(\mathbf{X}_{\text{a}}, \mathbf{X}_{\text{w}}, \mathbf{x}_{\text{cf}})^{m}\}_{m=1}^{M}$, of $M$ associated examples, where ``a'', ``w'', and ``cf'' are indices that associate the variables with the encoders. $\mathbf{X}_{\text{a}}^{m}\in\mathbb{R}^{T_{\text{a}}\times F_{\text{a}}}$ is a sequence of $T_{\text{a}}$ vectors of $F_{\text{a}}$ features of music audio signals, $\mathbf{X}_{\text{w}}^{m}\in\mathbb{R}^{T_{\text{w}}\times F_{\text{w}}}$ is a sequence of $T_{\text{w}}$ word embeddings of the musical genres assigned to $\mathbf{X}_{\text{a}}^{m}$ with $F_{\text{w}}$ features, and $\mathbf{x}_{\text{cf}}^{m}\in\mathbb{R}_{\geq0}^{1\times F_{\text{cf}}}$ is a vector of $F_{\text{cf}}$ features correlating $\mathbf{X}_{\text{w}}^{m}$ with a human created playlist.
By the encoders we obtain three latent representations and their information is mutually aligned using three contrastive losses between associated and non-associated examples. By the joint minimization of the losses, we obtain the optimized $e_{\text{a}}^{\star}$, later used for calculating embeddings of music signals (see Figure~\ref{fig:arch}).
\subsection{Obtaining the latent representations}
The audio encoder $e_{\text{a}}$ consists of $Z$ stacked 2D-CNN blocks, $\text{2DCNN}_{z}$, and a feed-forward block, FFB. Each $\text{2DCNN}_{z}$ consists of a 2D convolutional neural network ($\text{CNN}_{z}$) with a square kernel of size $K_{z}$ and unit stride, a batch normalization process (BN), a rectified linear unit (ReLU), and a pooling operation (\text{PO}). The FFB consists of a feed-forward layer, $\text{FF}_{\text{a1}}$, another BN process, a ReLU, a dropout with probability $p$, another feed-forward layer, $\text{FF}_{\text{a2}}$, and a layer normalization (LN) process. $e_{\text{a}}$ takes as an input $\mathbf{X}_{\text{a}}^{m}$ and the $Z$ 2D-CNN blocks and the feed-forward block process the input in a serial way. The output of $e_{\text{a}}$ is the learned representation $\boldsymbol{\phi}_{\text{a}}^{m}=e_{\text{a}}(\mathbf{X}_{\text{a}}^{m})$, computed as
\begin{align}
    &\mathbf{H}_{z}^{m} = \text{2DCNN}_{z}(\mathbf{H}_{z-1}^{m})\text{, and}\\
    &\boldsymbol{\phi}_{\text{a}}^{m} = \text{FFB}(\mathbf{H}_{Z}^{m})\text{, where}\\
    &\text{2DCNN}_{z}(u) = (\text{PO}\circ\text{ReLU}\circ\text{BN}\circ\text{CNN}_{z})(u)\text{,}\\
    &\text{FFB}(u) = (\text{LN}\circ\text{FF}_{\text{a2}}\circ\text{DP}\circ\text{ReLU}\circ\text{BN}\circ\text{FF}_{\text{a1}})(u)\text{, }
\end{align}
\noindent
$\mathbf{H}_{z}^{m}\in\mathbb{R}^{T''\times F''}$, $\mathbf{H}_{0}^{m} = \textbf{X}_{\text{a}}^{m}$, $\circ$ is the function composition symbol, i.e. $(f\circ g)(x)=f(g(x))$, and values of $T''$ and $F''$ depend on the hyper-parameters of $\text{CNN}_{z}$. 

The encoder $e_{\text{w}}(\cdot)$ is the genre encoder and consists of a self-attention (SA) over the input sequence, a feed-forward layer ($\text{FF}_{\text{w}}$), DP with probability $p$, an LN process, and a skip connection between the input of the feed-forward layer and its output. $e_{\text{w}}(\cdot)$ is after the self-attention mechanism employed in the Transformer model~\cite{vaswani2017attention}, and is used to learn a contextual embedding of its input, similarly to~\cite{favory2020learning}. 
Each musical genre associated with $\mathbf{X}_{\text{a}}^{m}$ is first one-hot encoded and then given as an input to the pre-optimized word embeddings model Word2Vec~\cite{mikolov:2013:iclr}. The output of Word2Vec is $\mathbf{X}_{\text{w}}^{m}$, which is then given as an input to $e_{\text{w}}(\cdot)$. The output of $e_{\text{w}}(\cdot)$ is the vector $\boldsymbol{\phi}_{\text{w}}^{m}=e_{\text{w}}(\mathbf{X}_{\text{w}}^{m})$, containing the contextual embedding of $\mathbf{X}_{\text{w}}^{m}$ and calculated as
\begin{align}
    \mathbf{V}'^{m} &= \text{SF}(\mathbf{X}_{\text{w}}^{m})\text{,}\\
    \mathbf{V}^{m} &= \mathbf{V}'^{m} + (\text{DP}\circ\text{FF}_{\text{w}})(\mathbf{V}'^{m})\text{, and}\\
    \boldsymbol{\phi}_{\text{w}}^{m} &= \text{LN}(\sum\limits_{i=1}^{T_{\text{w}}}\mathbf{V}_{i}^{m})\text{,}
\end{align}
\noindent
where $\mathbf{V}'^{m},\,\mathbf{V}^{m}\in\mathbb{R}^{T_{\text{w}}\times F'_{\text{w}}}$. 

The third encoder, $e_{\text{cf}}(\cdot)$, is the playlist association encoder and consists of a feed-forward block, similar to $e_{\text{a}}(\cdot)$, Specifically, $e_{\text{cf}}(\cdot)$ consists of a feed-forward layer, $\text{FF}_{\text{cf1}}$, a ReLU, a dropout process with probability $p$, another feed-forward layer, $\text{FF}_{\text{cf2}}$, and a LN process. The input to $e_{\text{cf}}(\cdot)$ is a vector, $\mathbf{x}_{\text{cf}}^{m}$, obtained by a collaborative filtering (CF) process, using $M_{\text{pl}}$ playlists created by humans.

The CF process gets as input a binary matrix, 
$\mathbf{B}_{\text{cf}}\in\{0, 1\}^{M\times M_{\text{pl}}}$, that indicates which songs are
included on which playlist, where $M_{\text{pl}}$ is the amount of playlists. Then, we minimize the WARP loss (Weighted Approximate-Rank Pairwise loss) using SGD and the sampling technique defined in~\cite{weston2011wsabie}, to approximate ranks between playlists and songs efficiently. CF outputs the matrices $\mathbf{X}_{\text{cf}}\in\mathbb{R}_{\geq0}^{M\times F_{\text{cf}}}$ and  $\mathbf{Q}_{\text{cf}}\in\mathbb{R}_{\geq0}^{F_{\text{cf}}\times M_{\text{pl}}}$, where $\mathbf{B}_{\text{cf}} \simeq \mathbf{X}_{\text{cf}}\cdot \mathbf{Q}_{\text{cf}}$. We use each row of $\mathbf{X}_{\text{cf}}$ as the vector $\mathbf{x}_{\text{cf}}^{m}$. We employ $e_{\text{cf}}(\cdot)$ to process the $\mathbf{x}_{\text{cf}}^{m}$, by providing a representation of $\mathbf{x}_{\text{cf}}^{m}$ that is learned specifically for the alignment process that our method tries to achieve. 
The output of $e_{\text{cf}}(\cdot)$ is the vector $\boldsymbol{\phi}_{\text{cf}}^{m}=e_{\text{cf}}(\mathbf{x}_{\text{cf}}^{m})$, calculated as
\begin{equation}
    \boldsymbol{\phi}_{\text{cf}}^{m} = (\text{LN}\circ\text{FF}_{\text{cf2}}\circ\text{DP}\circ\text{ReLU}\circ\text{FF}_{\text{cf1}})(\mathbf{x}_{\text{cf}}^{m})\text{.}
\end{equation}
\subsection{Optimization and alignment of latent representations}
We jointly optimize all encoders using $\mathbb{D}$ and three contrastive losses. We expand previous approaches on audio representation learning using multi-modal alignment, by employing multiple cross-modal and single modal alignment processes. Specifically, we align $\boldsymbol{\phi}_{\text{a}}^{m}$ with $\boldsymbol{\phi}_{\text{w}}^{m}$ (audio-to-genre, A2G, alignment), $\boldsymbol{\phi}_{\text{a}}^{m}$ with $\boldsymbol{\phi}_{\text{cf}}^{m}$ (audio-to-playlist, A2P, alignment), and $\boldsymbol{\phi}_{\text{cf}}^{m}$ with $\boldsymbol{\phi}_{\text{cf}}^{m}$ (genre-to-playlist, G2P, alignment). 

We use A2G alignment so that $\boldsymbol{\phi}_{\text{a}}^{m}$ 
keeps information about musical genre. Additionally, we further enhance the information in $\boldsymbol{\phi}_{\text{a}}^{m}$ by the A2P alignment, which is targeted to allow $\boldsymbol{\phi}_{\text{a}}^{m}$ to have information about playlist associations. Finally, we employ G2P alignment, so that we keep genre and playlist related information tied up together and not let them degenerate to some representation that just helps to minimize the employed losses. 
Specifically, we use the contrastive loss between two paired examples, $\pmb{\psi}_{\alpha}$ and $\pmb{\psi}_{b}$, defined as~\cite{chen2020simple,favory2020coala}
\begin{align}
    \mathcal{L}_{\pmb{\psi}_{\alpha},\pmb{\psi}_{b}} = \sum\limits_{i=1}^{M}-\log&\frac{\Xi(\pmb{\psi}^{i}_{\alpha}, \pmb{\psi}^{i}_{b}, \tau)}{\sum\limits_{k=1}^{2M}\mathbbm{1}_{[k\neq i]}\Xi(\pmb{\psi}^{i}_{\alpha}, \pmb{\zeta}^{k}, \tau)}\text{, where}\\
    \Xi(\mathbf{a}, \mathbf{b}, \tau) &= \exp(\text{sim}(\mathbf{a}, \mathbf{b})\tau^{-1})\text{,}\\
    \text{sim}(\mathbf{a}, \mathbf{b}) &= \mathbf{a}^{\top}\mathbf{b}(||\mathbf{a}||\;||\mathbf{b}||)^{-1}\text{,}\\
    \pmb{\zeta}^{k}&=\begin{cases}
    \pmb{\psi}^{k}_{a}\text{, if }k\leq M\\
    \pmb{\psi}^{k-M}_{b}\text{ else}
    \end{cases}\text{, }
\end{align}  
\noindent
$\mathbbm{1}_{\text{A}}$ is the indicator function with $\mathbbm{1}_{\text{A}} = 1$ iff A else 0, and $\tau$ is a temperature hyper-parameter.

We identify $\mathcal{L}_{\text{A2G}}=\mathcal{L}_{\boldsymbol{\phi}_{\text{a}},\boldsymbol{\phi}_{\text{w}}} + \mathcal{L}_{\boldsymbol{\phi}_{\text{w}},\boldsymbol{\phi}_{\text{a}}}$ as the loss for A2G alignment, $\mathcal{L}_{\text{A2P}}=\mathcal{L}_{\boldsymbol{\phi}_{\text{a}},\boldsymbol{\phi}_{\text{cf}}} + \mathcal{L}_{\boldsymbol{\phi}_{\text{cf}},\boldsymbol{\phi}_{\text{a}}}$ as the loss for A2P alignment, and $\mathcal{L}_{\text{G2P}}=\mathcal{L}_{\boldsymbol{\phi}_{\text{w}},\boldsymbol{\phi}_{\text{cf}}} + \mathcal{L}_{\boldsymbol{\phi}_{\text{cf}},\boldsymbol{\phi}_{\text{w}}}$ as the loss for G2P alignment. We optimize all of our encoders, obtaining $e_{\text{a}}^{\star}$, by minimizing the
\begin{equation}
    \mathcal{L}_{\text{tot}} = \lambda_{\text{A2G}}\mathcal{L}_{\text{A2G}} + \lambda_{\text{A2P}}\mathcal{L}_{\text{A2P}} + \lambda_{\text{G2P}}\mathcal{L}_{\text{G2P}}\text{,}
\end{equation}\noindent
where $\lambda_{\cdot}$ are different hyper-parameters used as weighting factors for the losses. 

\section{Evaluation}\label{sec:tasks}
To evaluate our method, we employ Melon Playlist Dataset~\cite{ferraro2021melon} as $\mathbb{D}$, in order to obtain $e_{\text{a}}$. Then, we assess the learned representations by $e_{\text{a}}$ applying it in different downstream tasks. Specifically, we focus on genre classification, audio-tagging, and automatic playlist continuation. For each of the tasks, we employ $e_{\text{a}}$ as audio encoder, which will provide embeddings to a classifier, trained for the corresponding task.

We assess the benefit of the contribution of each of the encoders of our method, by comparing our method using three encoders ($\text{Contr}_{\text{CF-G}}$) with our method but using only $e_{\text{a}}$ and $e_{\text{w}}$ ($\text{Contr}_{\text{G}}$), and using only $e_{\text{a}}$ and $e_{\text{cf}}$ ($\text{Contr}_{\text{CF}}$). 
In addition, we compare the performance on each task using a baseline architecture that directly predicts the target information from the audio encoder. We refer to these methods as $\text{B-line}_{\text{G}}$ for the model trained with genre information, $\text{B-line}_{\text{CF}}$ for the model trained to predict CF information and $\text{B-line}_{\text{CF-G}}$ for the model trained to predict both types of information at the same time.


\subsection{Melon Playlist Dataset and audio features}\label{sec:dataset}

The dataset $\mathbb{D}$ used to train the models was originally collected Melon, a Korean music streaming service. The dataset consists of $M$=649,091 songs, represented by their mel-spectrograms, and $M_{\text{pl}}$=148,826 playlists. The number of unique genres asociated with the songs is 219. In order to train the model we split the songs of the dataset in train (80\%), validation (10\%) and test (10\%). The split was done applying a stratified approach \cite{sechidis2011stratification} in order to assure a similar distribution of example in all the sets for the genres associated to the songs.

The pre-computed mel-spectrograms provided in the dataset correspond to a range of 20 to 50 seconds 
with a resolution of $F_{\text{a}}=48$ mel-bands. Such reduced mel-bands resolution did not negatively affect the performance of the auto-tagging approaches in our previous study~\cite{ferraro2020low} and have a significantly lower quality of reconstructed audio which allows to avoid copyright issues. Following the previous work~\cite{won2020evaluation}, we randomly select sections the songs to train the audio encoder, using $T_{\text{a}}=256$. \footnote{We trained using Tesla V100-SXM2 GPU with 32 GB of memory, the training took 19 minutes per epoch approximately.}





\subsection{Parameters optimization}

Following the best performance in previous work \cite{won2020multimodal, won2020evaluation} the audio encoder use $Z$=7 layers and $K$=3. We conducted a preliminary evaluation to select the hyper-parameters of the models, comparing the loss in the validation and training set to prevent the models of overfitting. We defined the dimensions for CF representations to $F_{\text{cf}}$= 300 and genres representations $F_{\text{w}}$= 200 with $T_{\text{w}}<=$ 10 genres per song. From the same preliminary evaluation we defined the temperature $\tau$=0.1, batch size of 128, learning rate of 1e-4, dropout of 0.5 and the number of heads for self-attention of 4. We did not experiment with changing the weights $\lambda$ for the different losses and we used $\lambda_{\text{A2G}}=\lambda_{\text{A2P}}=\lambda_{\text{G2P}}=1$.

\subsection{Downstream tasks}


Once the models are trained with the Melon Playlist Dataset, we use the pre-trained models to generate an embedding from the audio of each song in the different datasets. Then, we use the generated embeddings and compare the performance for each particular task. In the following, we describe each downstream task and the dataset used. 


\vspace{5pt}
\textbf{Genre Classification}
We use the fault-filtered version of the GTZAN dataset \cite{tzanetakis2002musical,kereliuk2015deep} consisting of music excepts of 30 seconds, single-labeled using 10 classes and split in pre-computed sets of 443 songs for training and 290 for testing.
We train a multilayer perceptron (MLP) of one hidden layer of size 256 with ReLU activations, using the training set and compute its accuracy on the test set. 
In order to obtain an unbiased evaluation, we repeat this process 10 times and average the accuracies.
We consider each embedding frame of a track as a different training instance, and when inferring the genres, we apply a majority voting strategy.
We also include the performance of pre-trained embedding models taken from the literature~\cite{cramer2019look,pons2019musicnn,gemmeke2017audio}, using the results reported in \cite{pons2019musicnn}.

\vspace{5pt}
\textbf{Automatic Tagging}.
We rely on the MTG-Jamendo dataset \cite{bogdanov2019mtg} which 
contains over 55,000 full audio tracks multi-labeled using 195 different tags from \textit{genre}, \textit{instrument}, and \textit{mood/theme} categories.\footnote{\url{https://mtg.github.io/mtg-jamendo-dataset/}}
For this task, we train a MLP that takes our pre-trained audio embeddings as input. 
We compute the embedding of all the tracks by averaging their embeddings computed on non-overlapping frames with the mean statistic.
The model is composed of two hidden layers of size 128 and 64 with ReLU activations, it includes batch normalizations after each layer and a dropout regularization after the penultimate layer.
We use the validation sets for early stopping and we finally evaluate the performances on the test sets using ROC AUC.
These evaluations are done on the three separated category of tags, each of them uses its own split. 
We repeat the procedures 10 times and report the mean average.


\vspace{5pt}
\textbf{Playlist Continuation}.
We make use of the playlists from the Melon Playlist Dataset that contain at least one track in our test set (not used when training our embedding model).
This provides 104,410 playlists, for the which we aim at providing 100 continuation tracks.
We compute the embedding of all the tracks by averaging their embeddings computed on non-overlapping frames with the mean statistic.
Then, for each track in a playlist, we compute the 100 most similar tracks, among the ones from the test set. 
These tracks are obtained using
the cosine similarity in the embedding space.\footnote{Similarity searches are computed using Annoy (\url{https://github.com/spotify/annoy}) based on Approximate Nearest Neighbors and angular distance \cite{dasgupta2008random}.}
Among all the retrieved similar tracks for a playlist, we finally select the 100 most repeated ones.
We compare these to the ground truth using 
normalized Discounted Cumulative Gain (nDCG) and Mean Average Precision (MAP) \cite{ricci}, which are commonly used to evaluate the performance of music recommendation systems.
These ranking metrics evaluate the order of the items for each playlist returned by the prediction. They return a higher score for a given playlist if the predicted ranked list contains items in the test set closer to the top. 

\section{Results}\label{sec:result}



\begin{table}[t]
\centering
\caption{GTZAN results}
\label{table}
\footnotesize
\setlength{\tabcolsep}{3pt}
\begin{tabular}{lc}
\hline
Model & Mean Accuracy $\pm$ STD \\
\hline
$\text{B-line}_{\text{G}}$  &  63.28 $\pm$ 1.19 \\
$\text{B-line}_{\text{CF}}$  & 57.12 $\pm$ 1.82\\
$\text{B-line}_{\text{CF-G}}$   & 64.35 $\pm$ 1.10 \\
$\text{Contr}_{\text{G}}$  &  \textbf{76.78} $\pm$ 1.22 \\
$\text{Contr}_{\text{CF}}$ & 67.12 $\pm$ 0.94 \\
$\text{Contr}_{\text{CF-G}}$ & 75.29 $\pm$ 1.32 \\
\hline
VGGish Audioset \cite{gemmeke2017audio}  & 77.58 \\
OpenL3 Audioset \cite{cramer2019look} & 74.65 \\
musicnn MSD \cite{pons2019musicnn}  &77.24 \\
\hline
\end{tabular} 
\label{tab:gtz}
\end{table}

Focusing on \textbf{genre classification}, the results in Table~\ref{tab:gtz} show that the performance of the audio embedding when trained using the contrastive loss is always higher than using the models trained directly to predict the modality information ($\text{B-line}$). The best performance is obtained with $\text{Contr}_{\text{G}}$ with a similar result to when also considering CF information when training the embedding model ($\text{Contr}_{\text{CF-G}}$).
We also see that the performances of the $\text{Contr}_{\text{G}}$ model are comparable with state-of-the-art pre-trained embeddings (VGGish audioset) \cite{gemmeke2017audio, pons2019musicnn}.
This is particularly interesting since a large percentage of the Melon Playlist Dataset consists of korean music, which can be different from popular western music from the GTZAN collection.


\begin{table}[t]
\centering
\caption{Automatic tagging results}
\label{table}
\footnotesize
\begin{tabular}{lccc}
 \hline
   & \multicolumn{3}{c}{ROC AUC $\pm$ STD}\\
  Model & Genre & Mood & Instrument\\
\hline
$\text{B-line}_{\text{G}}$ & 0.840 $\pm$ 0.004& 0.722 $\pm$ 0.004 & 0.781 $\pm$ 0.005\\
$\text{B-line}_{\text{CF}}$ & 0.836 $\pm$ 0.002 & 0.722 $\pm$ 0.003 & 0.770 $\pm$ 0.008\\
$\text{B-line}_{\text{CF-G}}$ & 0.845 $\pm$ 0.004 & 0.727 $\pm$ 0.006 & 0.785 $\pm$ 0.004\\
$\text{Contr}_{\text{G}}$ & \textbf{0.847} $\pm$ 0.004 & 0.732 $\pm$ 0.005 & \textbf{0.797} $\pm$ 0.005\\
$\text{Contr}_{\text{CF}}$ & 0.845 $\pm$ 0.004 & 0.732 $\pm$ 0.004 & 0.793 $\pm$ 0.007\\
$\text{Contr}_{\text{CF-G}}$ & 0.843 $\pm$ 0.004 & \textbf{0.733} $\pm$ 0.005 & 0.791 $\pm$ 0.006\\
\hline
\end{tabular} 
\label{tab3}
\end{table}

\vspace{5pt}
\textbf{Automatic Tagging}. From the results in Table~\ref{tab3} we see that the methods based on contrastive learning outperform the baselines in almost all the cases. 
The best results for the instrument and genre tags is obtained with the $\text{Contr}_{\text{G}}$ model. For the mood tags the best performance is achieved with $\text{Contr}_{\text{CF-G}}$, which takes advantage of the information in the playlists and the genre annotations.


\begin{table}[t]
\centering
\caption{Playlist generation results}
\label{table}
\footnotesize
\setlength{\tabcolsep}{3pt}
\begin{tabular}{lcc}
\hline
Model & NDCG@100 & MAP@100 \\
\hline
random & 0.0005& 0.0001\\
$\text{B-line}_{\text{G}}$ & 0.0044 & 0.0007  \\
$\text{B-line}_{\text{CF}}$ & 0.0035 & 0.0007  \\
$\text{B-line}_{\text{CF-G}}$ & 0.0042 & 0.0008  \\
$\text{Contr}_{\text{G}}$ & 0.0074 & 0.0016 \\
$\text{Contr}_{\text{CF}}$  & 0.0076 & 0.0017 \\
$\text{Contr}_{\text{CF-G}}$ & \textbf{0.0085} & \textbf{0.0020}\\
\hline
\end{tabular} 
\label{tab2}
\end{table}

\vspace{5pt}
The results for the task of \textbf{Automatic playlist continuation} follow the same trend of the other tasks. The models trained using the contrastive loss perform better than the baselines trained directly to predict the genres or the CF representation. 
The best performance is obtained with the $\text{Contr}_{\text{CF-G}}$ model, which combines genre and CF information.  

 
\section{Conclusions}\label{sec:conclusions}

In this work, we propose a method for learning an audio representation, by combining multiple sources of information related to the music using contrastive learning. 
We evaluate the method by pre-traing the model using information from the Melon Playlist Dataset and we compare the performance in three downstream tasks in the music domain (genre classification, automatic tagging, and automatic playlist continuation). 
We see that using contrastive learning allows us to reach higher performance than using the models trained directly to predict the genre or the collaborative filtering information. This indicates that contrastive learning is effective at learning simultaneously from heterogeneous information, enabling us to improve the overall performance across different tasks. 

The dataset used for training our embedding model offers additional types of information that we did not use. They include title, playlist tags and authors, as well as other metadata of the tracks. As future work, we propose incorporating this playlist-level information which will require an additional level of abstraction to our architecture. 
%
%
%




\bibliographystyle{IEEEtran}
\bibliography{references}
\end{document}